\documentstyle[11pt,newpasp,twoside,epsfig]{article}
\def\gtrapprox{\;\lower 0.5ex\hbox{$\buildrel >\over \sim\ $}}
\def\lessapprox{\;\lower 0.5ex\hbox{$\buildrel < \over \sim\ $}}          
\def\Msun  {${\rm M}_\odot$}
\def\deg   {$^\circ$}

\def\HI    {H{$\rm\scriptstyle I$}}

\def\kms   {\ km s$^{-1}$}

\def\edcomment#1{\iffalse\marginpar{\raggedright\sl#1\/}\else\relax\fi}
\reversemarginpar

\begin{document}
\title{Neutral Hydrogen and the Missing Satellites of the Local Group}

\author{M.E. Putman}
\affil{CASA, University of Colorado, Boulder, CO 80309-0389; Hubble Fellow}
\author{Ben Moore}
\affil{Department of Physics, Durham University, UK}

\begin{abstract}
We present a comparison of the leftover satellites at z=0 in a cold dark matter
dominated simulation of the formation of the Local Group to
the distribution of observed neutral hydrogen high-velocity clouds
and compact high-velocity clouds.
The $\sim$2000 leftover satellites in the simulation have dark matter masses 
which range
between 0.5 to 10 $\times\ 10^9$ \Msun, sizes
between 3 to 30 kpc, and distances between 100 kpc and 2 Mpc.  The dark
matter halos show a clear bias
in their distribution towards M31 and to a lesser extent towards
the Local Group anti-barycenter.
If the Local Group halos contain $\sim$1\% of their dark matter mass in neutral hydrogen
they should have been easily detected by the current \HI\ surveys.  The
only \HI\ objects detected with the potential to be the Local Group halos are the
high-velocity clouds.
Here the spatial, kinematic, and \HI\ flux properties of the clouds and dark matter halos 
are compared.  Several different subsets of halos which may be more
likely to contain neutral hydrogen are investigated, and the HVCs are found
to have some similar properties to those halos within 500 kpc of the Galaxy
and those halos with dark matter 
masses $> 2 \times\ 10^8$ \Msun.  The compact high-velocity clouds do
not show similar properties to the halos. 

\end{abstract}

\section{Introduction}

The distribution of neutral hydrogen in the universe is largely traced by 
the Ly$\alpha$ forest, Ly-limit systems, damped Ly$\alpha$ systems and
galaxies, but the relationship between these tracers remains unknown.
The predicted existence of a large number of individual dark 
matter halos in groups of galaxies by the cold dark matter (CDM) models 
of the formation of galaxies and clusters may provide a clue to the 
link between these features.   
If the excess satellites/halos exist, do they trace the 
\HI\ distribution of the group and explain the abundance and location of
the absorbers?  All-sky HI surveys have detected only one type of object
which could be the abundance of dark matter halos scattered throughout the
Local Group.  These are the high-velocity clouds (HVCs).

Using a standard CDM model of the formation of the Local Group (Moore et al. 2001), 
we compare the properties of the cold dark matter halos with those of the
high-velocity clouds and compact high-velocity clouds (CHVCs) to determine if these objects may solve
the ``missing satellite'' problem.  
If the halos trace the \HI\ distribution of a cluster, we are also able to place
strong constraints on the fraction of neutral hydrogen associated with 
the dark matter.

\section{Observations}

The high-velocity cloud \HI\ properties come from the $\sim35^{\prime}$ spatial 
resolution catalog of Wakker (priv comm). 
The Wakker catalog is adapted from the catalog of Wakker \& van Woerden (1991)
and uses data from the Dwingeloo Telescope
in the north (Hulsbosch \& Wakker 1988) and the IAR 30m in the south
(Morras et al. 2000).  This HVC catalog was used because it is from a
fairly homogeneous 
resolution and sensitivity all-sky HVC dataset and uses a single catalog method. 
The data have a 5$\sigma$ brightness temperature sensitivity of
$\sim$0.05 K when it is smoothed kinematically.
In this catalog of 626 clouds, the large complexes such as Complex C
and the Magellanic Stream are classified as one object, leading to
a distinct difference in the number of individual objects between this catalog 
and the southern HVC catalog 
of Putman et al. (2001).  The Putman et al. catalog (described below) was designed to catalog
smaller objects and only merges clouds if the brightest enclosing contour is 
$> 0.4$ T$_{max}$.  This catalog has 1956 clouds in it and the flux
distribution has a significantly steeper slope.  See Putman et al. for the
distribution plots of this catalog.

We also investigated the properties of the compact high-velocity clouds (CHVCs).  CHVCs
were originally defined by Braun \& Burton (1999) as isolated clouds with 
diameters less than 2\deg\ (based on the 50\% of peak N$_{HI}$ contour).
This definition was slightly revised in the higher resolution and sensitivity 
Putman et al. catalog of CHVCs to use the 25\% of the peak N$_{HI}$ contour.
The HIPASS (HI Parkes All-Sky Survey; Barnes et al. 2001) data used in the Putman et al. catalog provide a marked
improvement in the detectability of CHVCs, so this catalog has been combined with the
predominantly northern catalog of Braun \& Burton, which was extracted
primarily from the Leiden-Dwingeloo Survey (LDS) data ($5\sigma \approx 0.1$K).
The total number of CHVCs is 215.
The Putman et al. catalog covers declinations less than +2\deg\
and uses HIPASS data reduced with
the {\sc minmed5} method (Putman 2000).
HIPASS has a spatial resolution of 15.5$^{\prime}$, a spectral resolution, after 
Hanning smoothing, of 26.4 km s$^{-1}$, and a 5$\sigma$ rms brightness 
temperature 
sensitivity of approximately 0.04 K, which corresponds to a $5\sigma$ 
column density sensitivity of $2.5 \times 10^{18}$ atoms cm$^{-2}$ 
for an HVC with a linewidth of 35\kms.  The HIPASS HVC data is complete
to fluxes of 2$-$4 Jy \kms, however the catalog may only be 100\%
complete to total fluxes of 25 Jy \kms\ (see Putman et al. 2001). 

The primary selection effect when cataloging HVCs and CHVCs is the gap in the
velocity distribution due to the presence of Galactic HI.  Clouds between
$\mid v_{LSR} \mid$ = 90 \kms\ are missing in all of the catalogs discussed above.
This can be translated to a missing $v_{LGSR}$ population with
the equation,  $\mid v_{LGSR}\mid$ = $v_{LSR}$ + 220sin($\ell$)cos($b$) - 
62cos($\ell$)cos($b$) + 40sin($\ell$)cos($b$) - 35sin($b$) (Braun \& Burton
1999).

\section{Simulation}

We analyze a binary pair of massive halos that form in a hierarchical universe
dominated by cold dark matter. The masses, separation and relative velocities
of the binary halos are close to those observed for the Milky Way and
Andromeda. The binary system was chosen from a large cosmological system such
that a nearby massive cluster similar to Virgo was present. The simulation is
described in detail in Moore et al. (2001). At $z=0$, each halo contains $\sim 2\times 10^6$
particles and is resolved to a distance of 0.1\% of the virial radius,
$r_{200}\approx 200$ kpc.  The peak circular velocities of the halos are 220
\kms\ and 200 \kms. These two halos are infalling for the first time at
$\approx 100$ \kms\ and their outer edges are separated by 700 kpc.  Over
2000 smaller dark matter
halos lie within the high-resolution region and mostly within the virial
radii of the two large halos that represent the Milky Way and Andromeda.  It
is these low mass dark matter halos that constitute the ``missing satellite''
problem for CDM models -- the Local Group galaxies make up just 1-2\% of the
expected number of satellites. 

\section{Results}

Figure 1 shows the spatial distributions of all of the Local Group dark
matter halos, 
the high-velocity clouds (without distance constraints; see below), and the compact high-velocity
clouds in Galactic coordinates.   The dark matter halos show
a clear bias in their distribution towards M31 and the Local Group
anti-barycenter region. This bias in the distribution is independent of the size
and mass of the halos, however, in terms of distance, it only appears 
beyond $\sim$500 kpc of the Milky Way (see Figure 2).
The distribution of cataloged HVCs shown in the middle panel of Figure 1, excludes those
HVCs which have distance determinations.  
Direct distance determinations have been made to several of the larger
high velocity complexes which place them within approximately 10 kpc
of the Galaxy (e.g. Complex A, M, and WE; Wakker 2001), and  
the Stream and Leading Arm originated
from the interaction of the Magellanic Clouds with the Milky Way and
are at distances within 100 kpc.
We can also exclude those complexes which are
bright in H$\alpha$ emission (e.g. Complex L, GCP (the Smith Cloud), 
and C), as this emission would not have been detected at
distances greater than 100 kpc (e.g. Bland--Hawthorn \& Maloney 1999), 
unless associated with an active star forming region.
There are excesses of HVCs in the same general
directions as the halos, but the extreme overabundance in the direction of M31
is not present.   
Furthermore, the distance determination
methods have not uniformly sampled the sky, and this bias in the
HVC distribution may disappear with future observations.
The CHVCs shown in the bottom panel of Figure 1 are clustered, but 
clearly show a different distribution to the Local Group halos.  

\begin{figure}
\plotfiddle{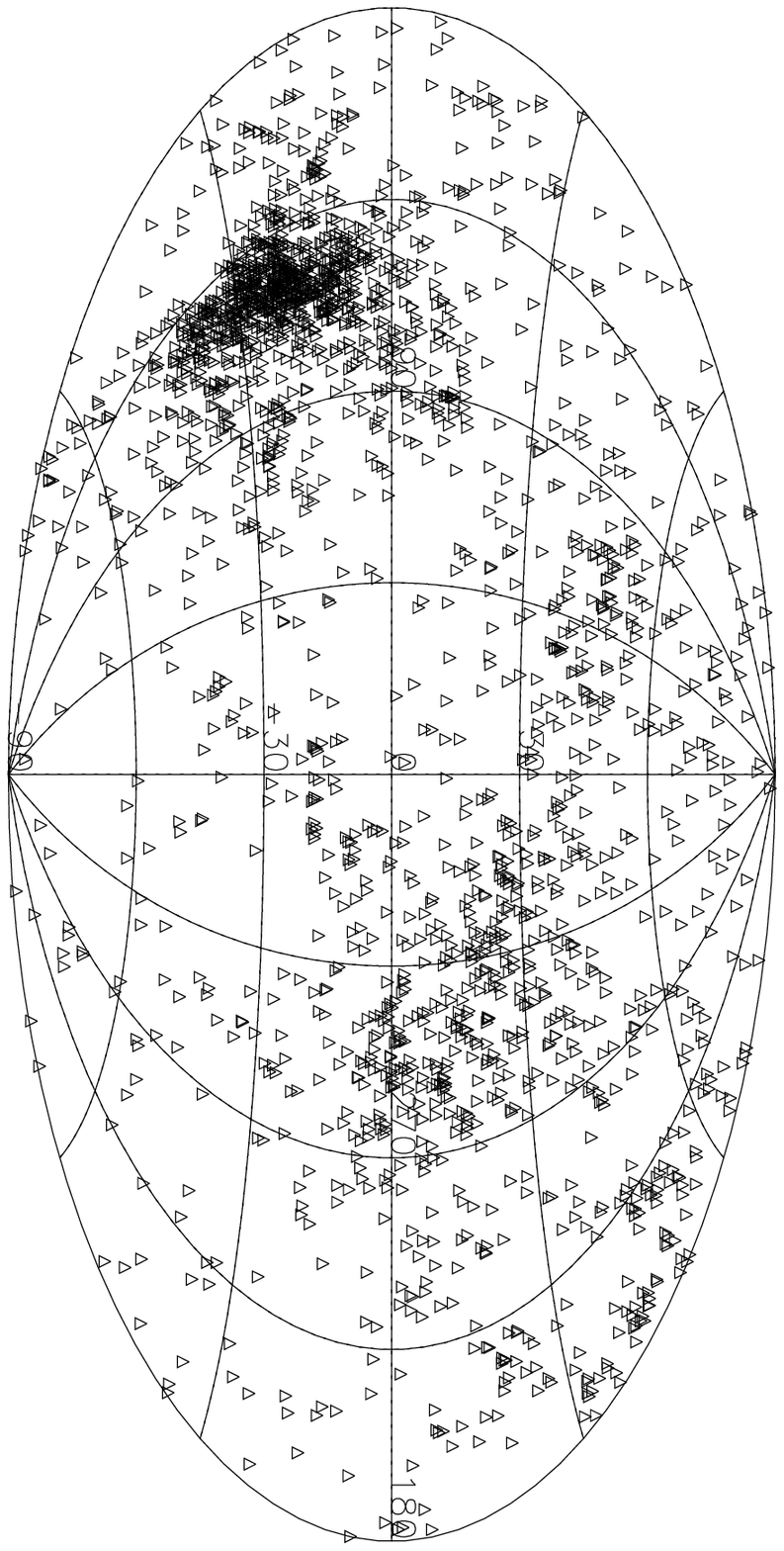}{2.4in}{90}{55}{55}{250}{-90}
\plotfiddle{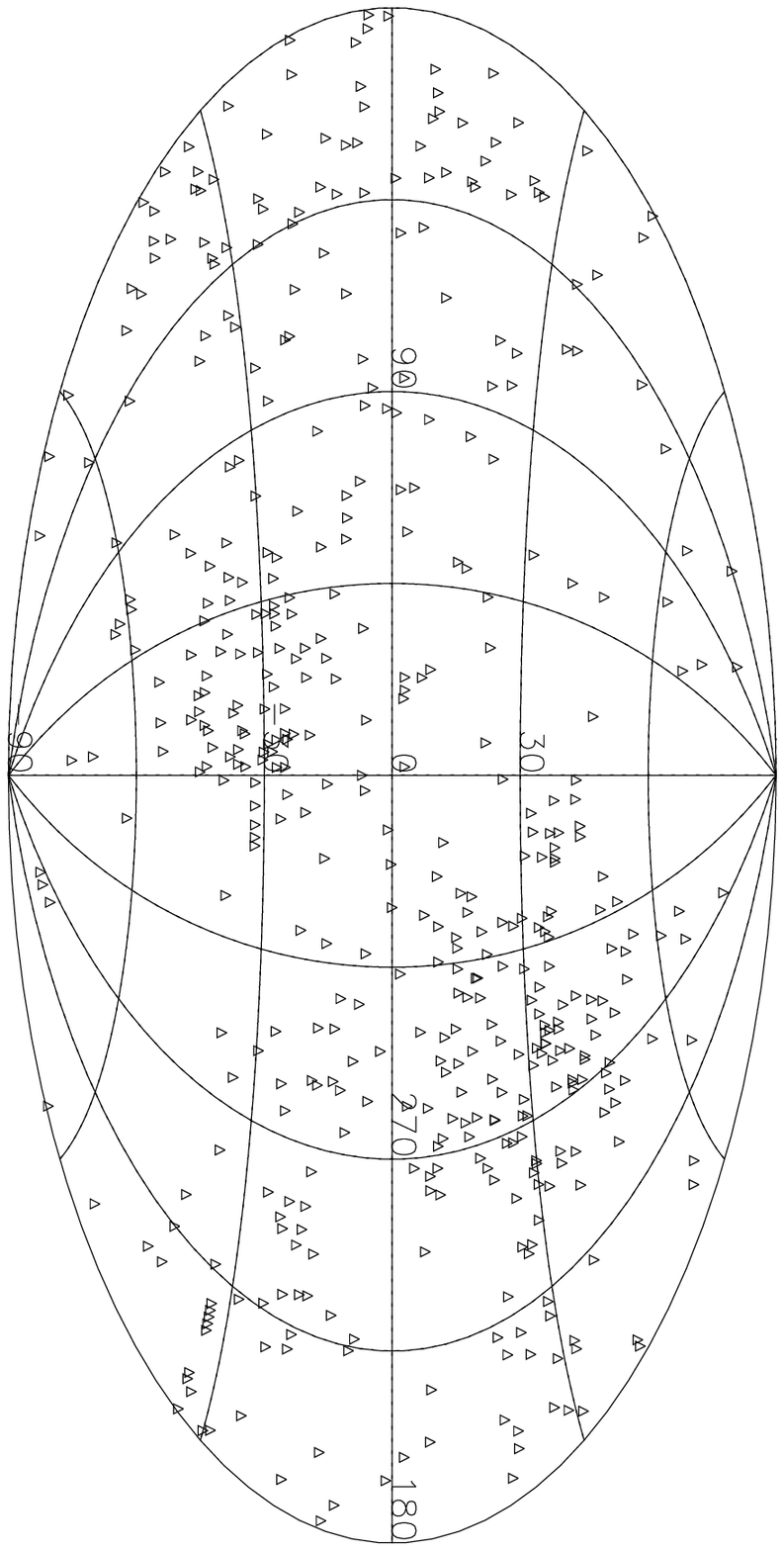}{2.4in}{90}{55}{55}{250}{-90}
\plotfiddle{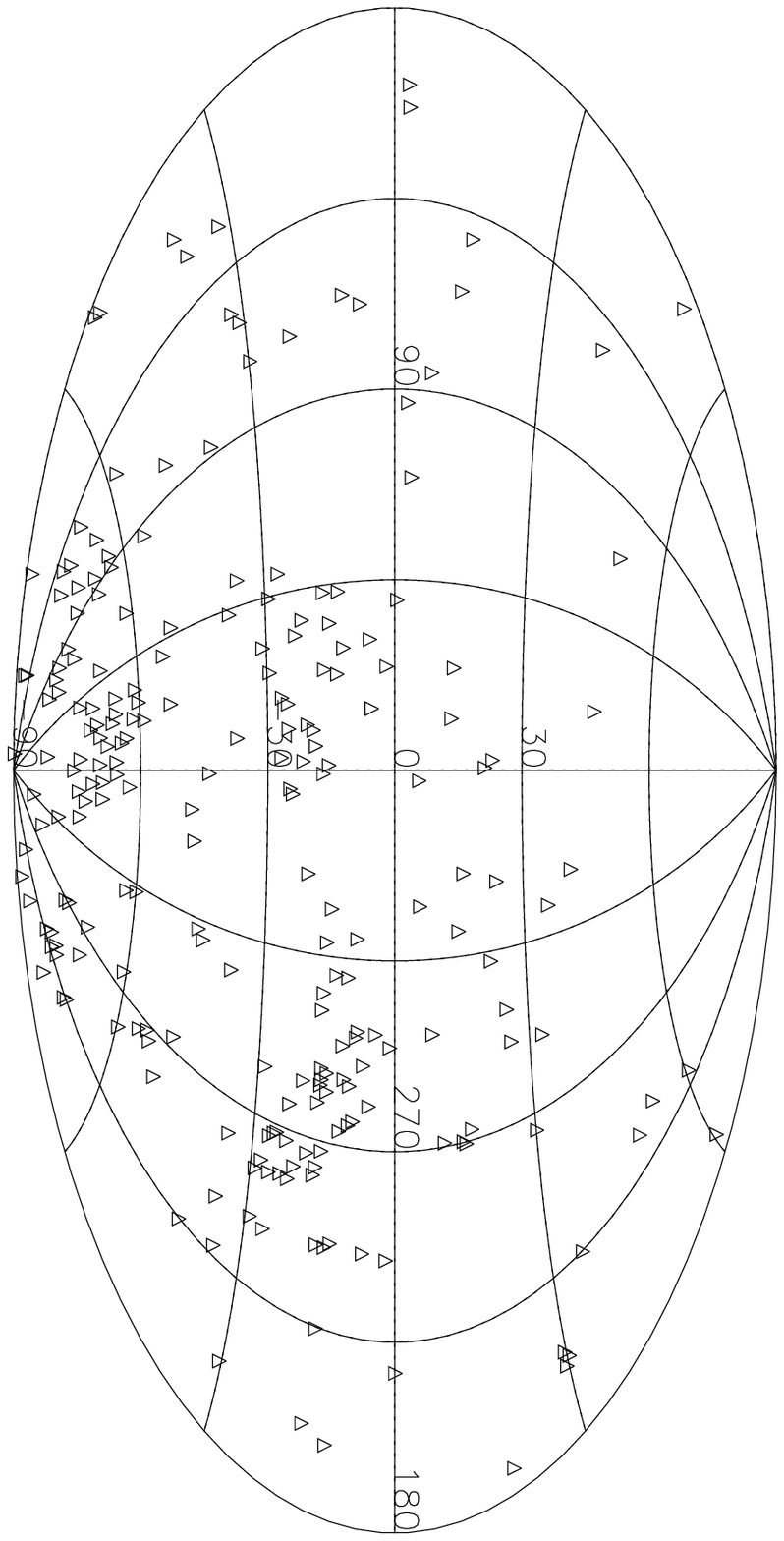}{2.4in}{90}{55}{55}{250}{-90}
\caption{The spatial distribution of all of the Local Group dark matter halos (top), high-velocity clouds
(middle; excluding those clouds with distance constraints),  and compact
high-velocity clouds (bottom) in Galactic coordinates.}
\end{figure}

\begin{figure}
\plotfiddle{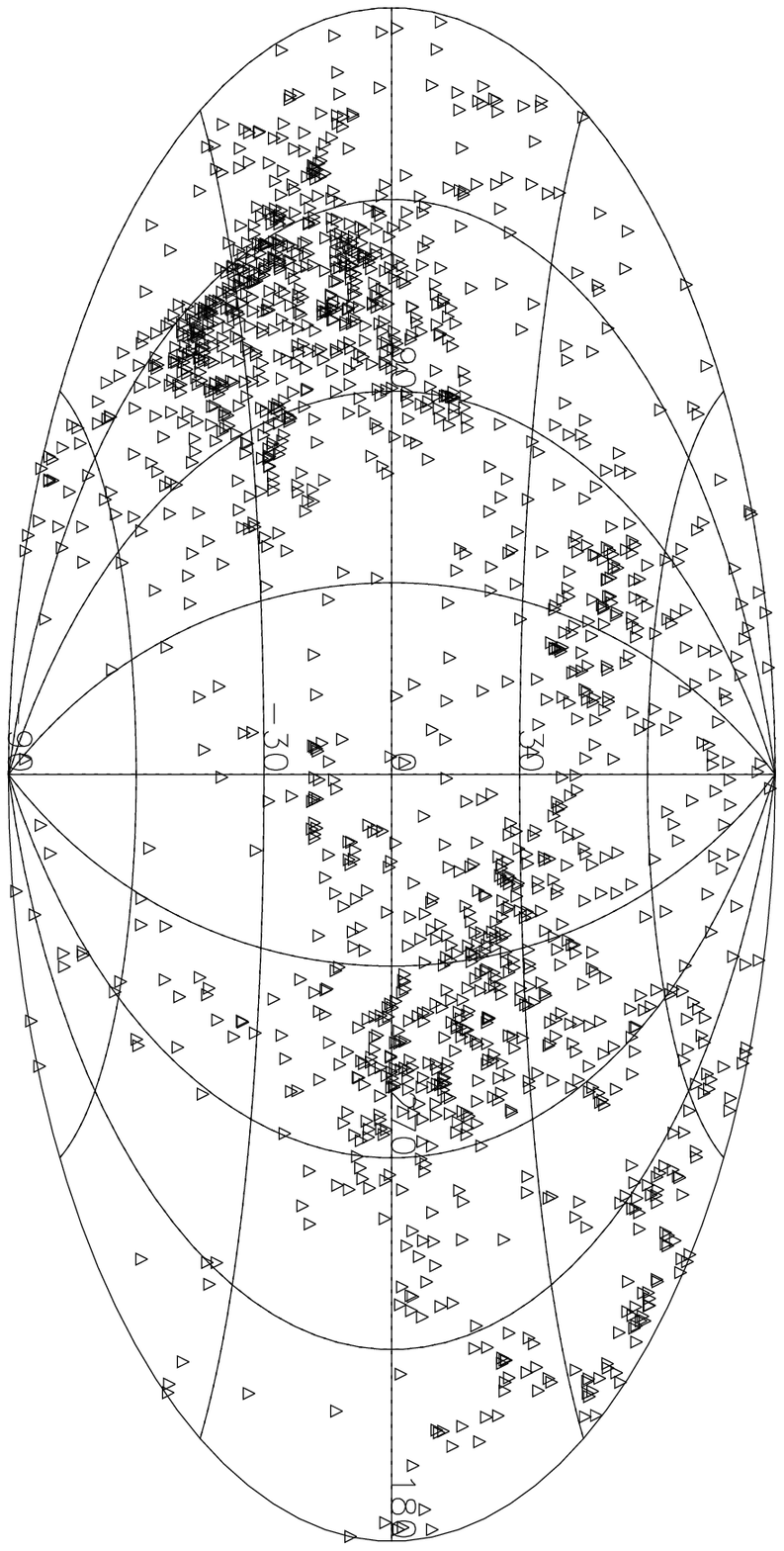}{2.4in}{90}{55}{55}{250}{-90}
\plotfiddle{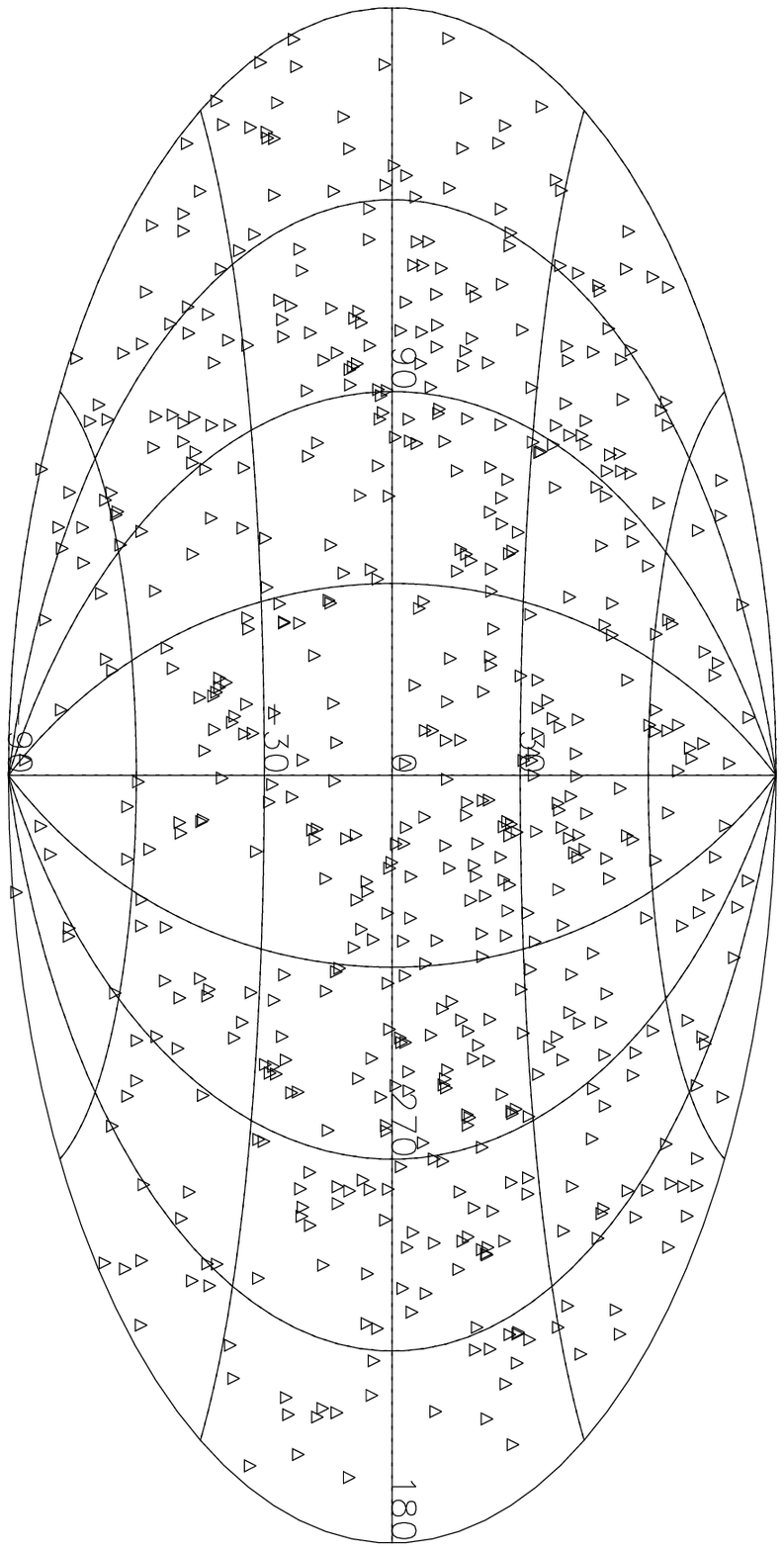}{2.4in}{90}{55}{55}{250}{-90}
\plotfiddle{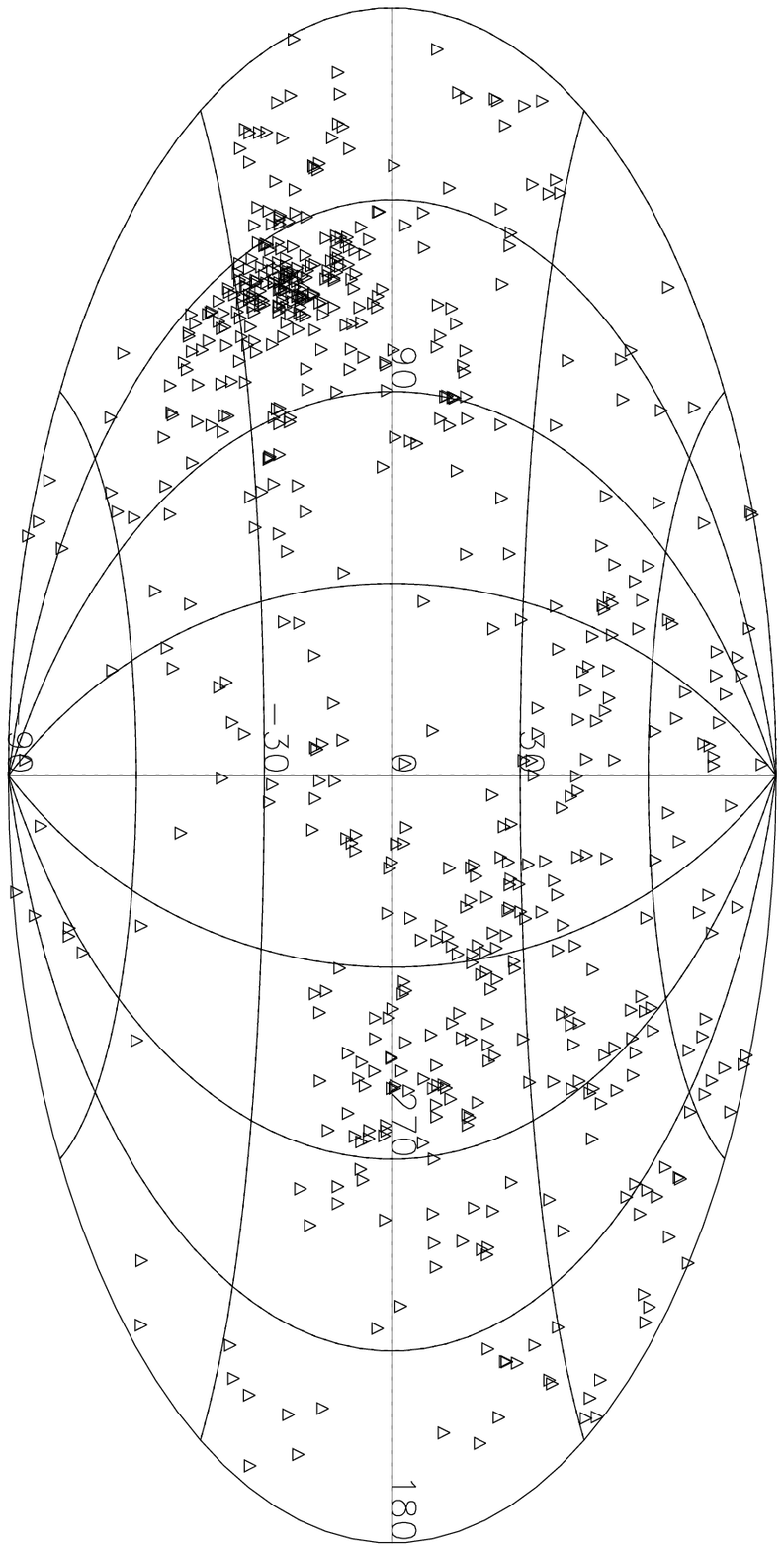}{2.4in}{90}{55}{55}{250}{-90}
\caption{The spatial distribution of the Local Group halos more than 300 kpc away from
the Galaxy and M31 (top), within 500 kpc of the Galaxy
(middle), and those halos with dark matter masses greater than $2 \times
10^8$ \Msun\ (bottom) in Galactic coordinates.}
\end{figure}

\begin{figure}
\plotfiddle{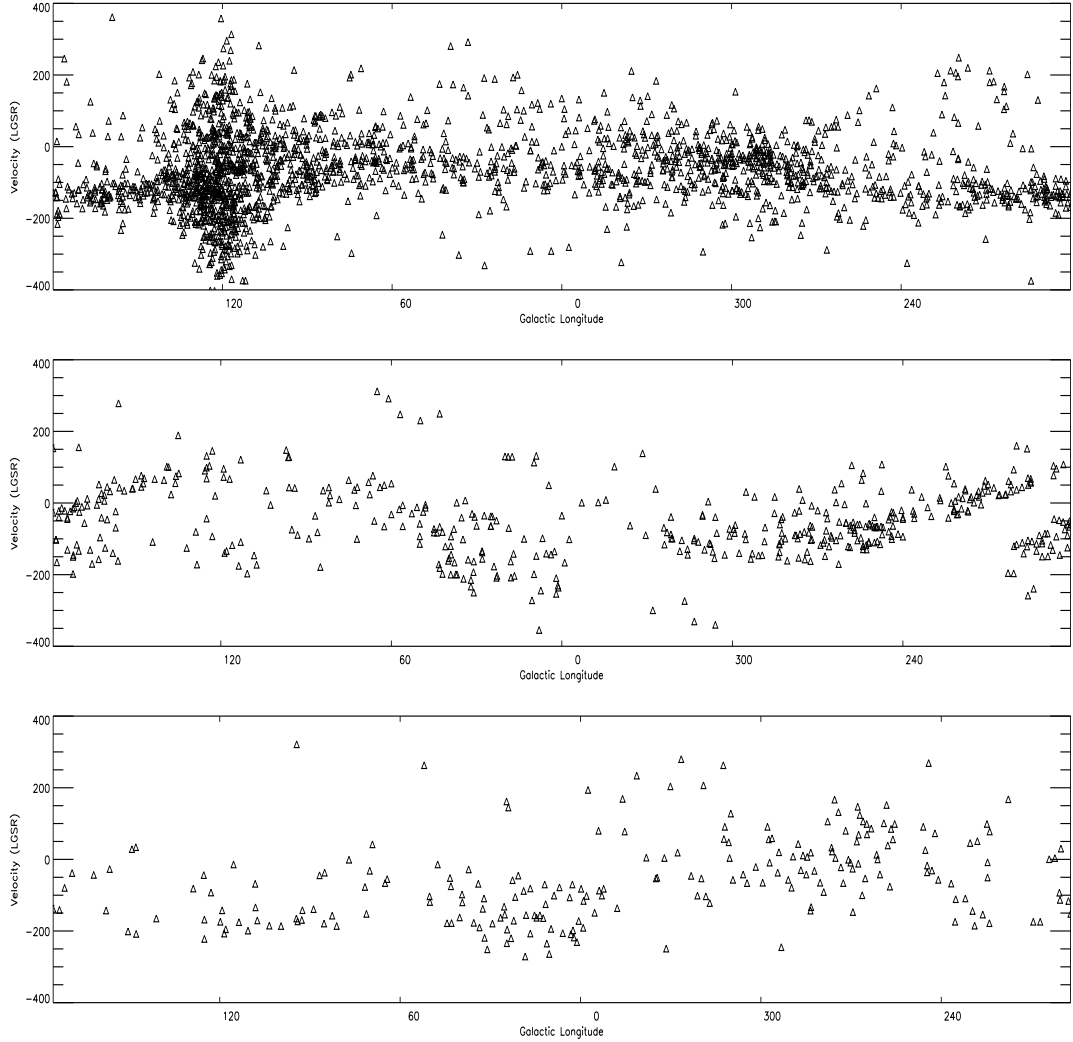}{5.5in}{90}{60}{80}{240}{-50}
\caption{The velocity distribution of the Local Group halos (top), HVCs
(middle; excluding those clouds with distance constraints) and CHVCs (bottom) in
terms of the Local Group Standard of Rest ($v_{\rm LGSR}$).}
\label{fig:vel}
\end{figure}

\begin{table}
\caption[]{Cloud and Halo Properties}
\label{tab:vel}
\begin{tabular}{rrc}
\tableline
\tableline
Object & \# & Mean $v_{\rm LGSR}$ \\
\tableline
HVCs (without distance constraints)& 469 & -49  \\
CHVCs (Compact HVCs) & 215 & -57 \\
All dark matter halos & 2135 & -66 \\
Halos outside of 300 kpc & 1530 & -69 \\
Halos within 500 kpc & 584 & -27 \\
Halos with M$_{DM} > 2 \times 10^8$ \Msun\ & 624 & -63 \\
\tableline
\end{tabular}
\end{table}

\begin{figure}
\plotfiddle{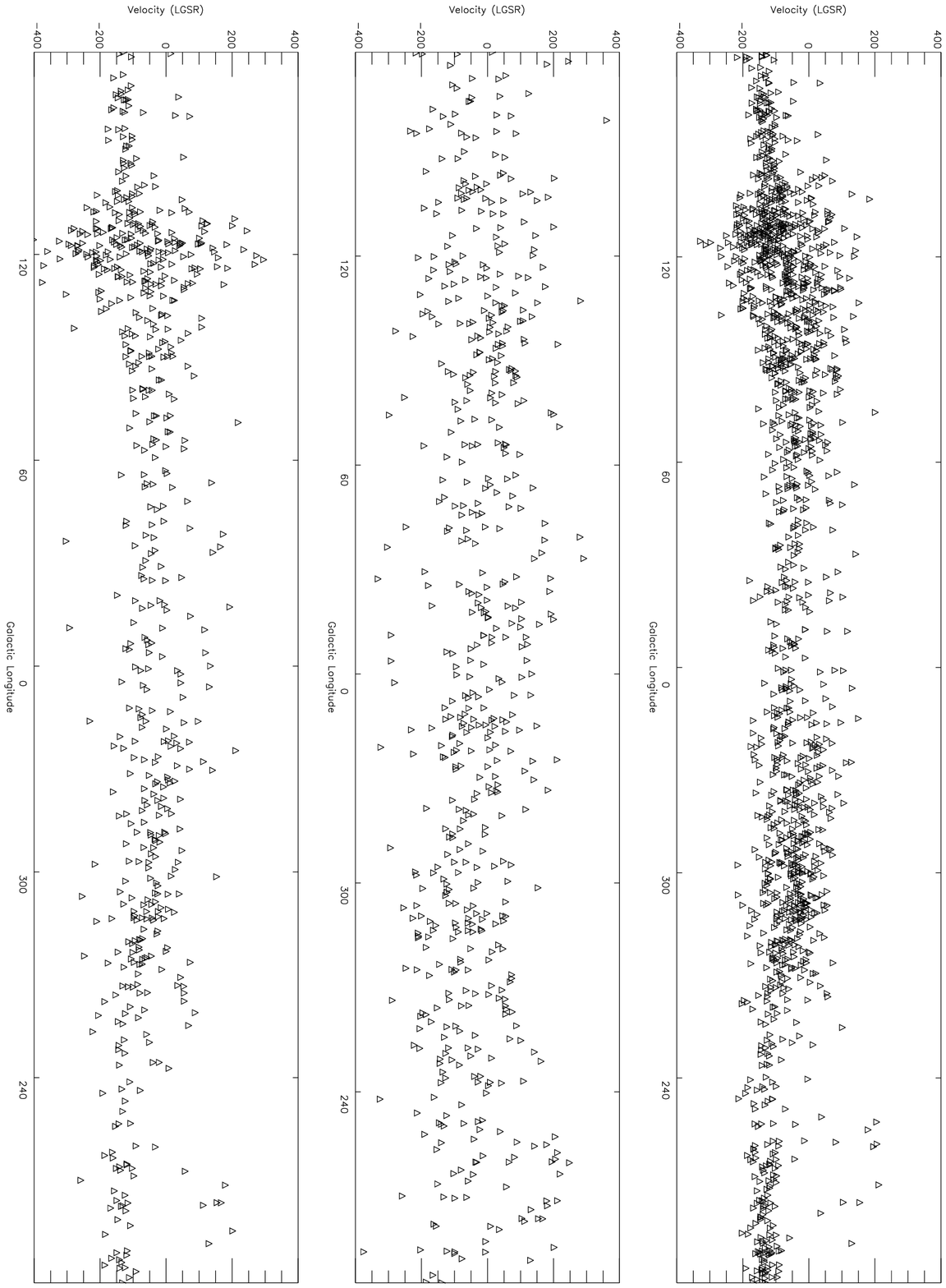}{5.5in}{90}{60}{80}{240}{-50}
\caption{The velocity distribution ($l$ vs. $v_{\rm LGSR}$) of the Local Group halos beyond 300 kpc
of both the Galaxy and M31 (top), at distances less
than 500 kpc from the Galaxy (middle) and with dark matter masses greater
than $2 \times 10^8$ \Msun\ (bottom).}
\label{fig:lgveldist} 
\end{figure}

Figure 2 shows the spatial distributions of only those Local Group halos at least 300 kpc from both
M31 and the Galaxy (top), within 500
 kpc of the Galaxy (middle), and those halos with dark matter masses $> 2 \times 
10^8$ \Msun.  The bias in the spatial distribution of the halos is 
not evident within 500 kpc of the Galaxy.  The beginning of an asymmetric
distribution is seen between 500 and 800 kpc in this simulation.
Since the simulation places the center of M31 at 1100 kpc, rather than the
observed distance of 700 kpc (Freedman \& Madore 1990), the
actual bias in the distribution could appear somewhat closer.  The asymmetric distribution of halos remains in both the $>$ 300 kpc 
and $> 2 \times 10^8$ \Msun\ cases, but the total number of objects is greatly decreased
(see Table 1).

The velocity distributions of all of the dark matter halos, HVCs without distance
constraints, and CHVCs are shown in terms of Galactic longitude ($l$) and Local Group Standard of
Rest velocity ($v_{\rm LGSR}$) in
Figure 3.  Their mean LGSR velocities are tabulated in Table 1.  There is no obvious position-velocity relationship
between the populations of clouds and the Local Group halos.
A wide range of velocities about M31 is shown in the
simulation, as is a clear bias for negative $v_{\rm LGSR}$'s (i.e. infall).
The HVC and CHVC velocity distributions are limited by confusion with
Galactic emission between $v_{LSR} \approx$ +/- 90~\kms.
This is evident in the HVC distribution shown in the middle panel of Figure
3, with the HVCs clustering about a sinusoidal gap in the 
distribution.
Both the HVCs and CHVCs show a preference for negative velocities
in the Local Group reference frame.
The CHVCs show a tendency towards more positive velocities at $l > 180$\deg\ and
negative velocities at $l < 180$\deg,
with the largest range of velocities about $l = 0$\deg. 
The panels of Fig. 4 show the velocity distributions of the selected groups of 
halos shown in Fig. 2.
The velocity distribution of the Local Group halos within 500 kpc is fairly uniform, 
as is the spatial distribution, with a mean $v_{LGSR}$ the same magnitude
more positive than the HVCs ($\sim 20$ \kms) as the other halo populations are
more negative.  The population of halos with M$_{DM} > 2 \times 
10^8$ \Msun\ appears as a thinned version of the velocity distribution of all of 
the halos.  Those halos over 300 kpc from M31 and the Galaxy have a
similar thinned distribution, but there is less scatter and less extreme velocities in the direction of M31.

\begin{figure}
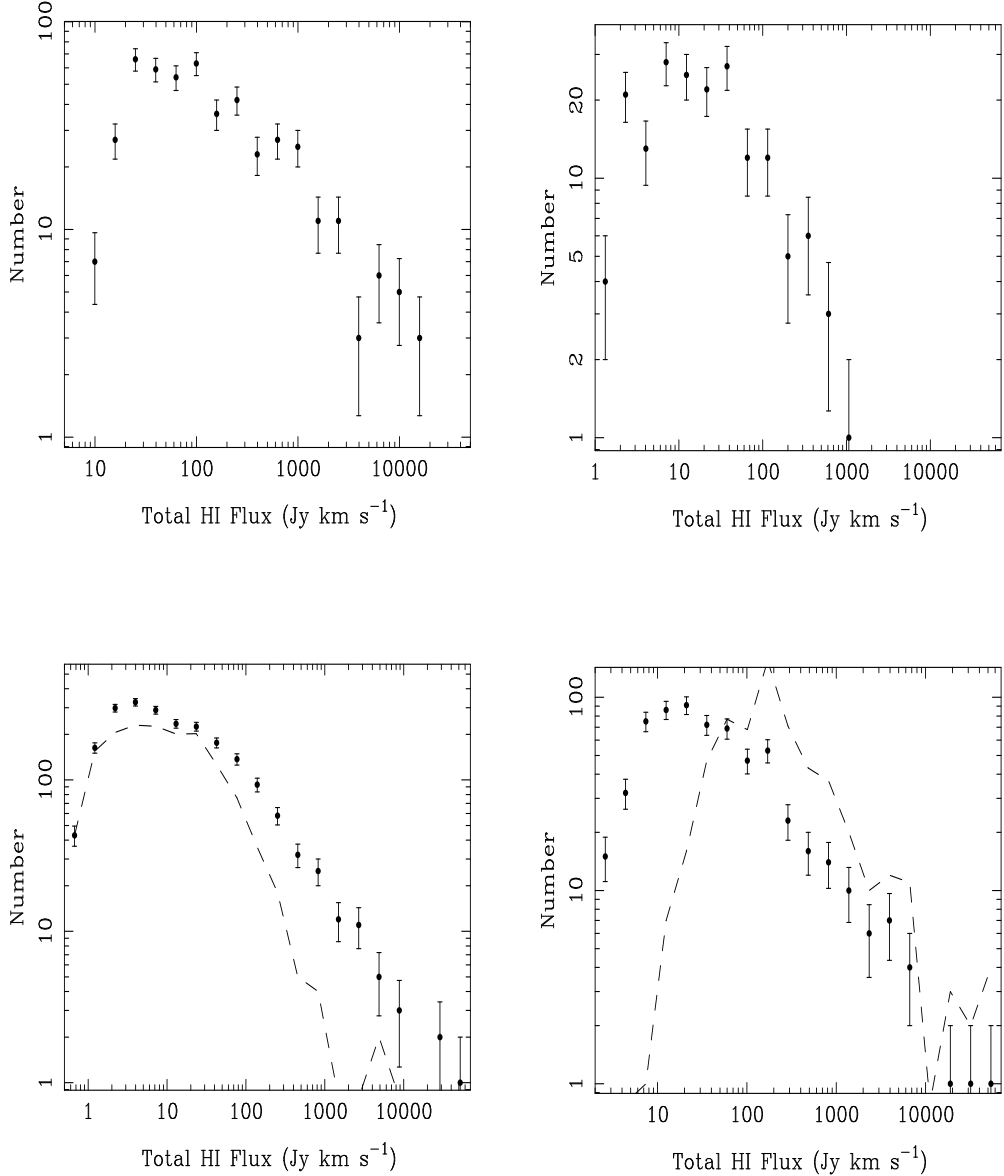

\plotfiddle{mputman_5a.ps}{1.5in}{0}{40}{50}{-220}{-210}
\plotfiddle{mputman_5b.ps}{1.5in}{0}{40}{50}{-20}{-90}
\plotfiddle{mputman_5c.ps}{1.5in}{0}{40}{50}{-220}{-210}
\plotfiddle{mputman_5d.ps}{1.5in}{0}{40}{50}{-20}{-90}
\caption{The \HI\ flux distribution of the HVCs (upper left; $f(F_{HI})
\propto F_{HI}^{-1.5}$), CHVCs (upper right; $f(F_{HI})
\propto F_{HI}^{-1.8}$), and the Local Group
halos if the halos currently contain 1\% of their
dark matter mass in neutral hydrogen. Note the different axes values on
these plots.  The bottom left panel
shows all of the dark matter halos (solid points; $f(F_{HI}) \propto F_{HI}^{-1.5}$) and those halos outside of 300 kpc from
both M31 and the Galaxy (dashed line; $f(F_{HI}) \propto F_{HI}^{-1.8}$).
The bottom right panel shows those halos with masses less than $2 \times 10^8$
\Msun (solid points; $f(F_{HI}) \propto F_{HI}^{-1.5}$) and within 500 kpc of the
Galaxy (dashed line; $f(F_{HI}) \propto F_{HI}^{-1.6}$).  The data is binned in equal intervals of
$\log(F_{HI})$ and the slopes are quoted in linear units.}
\label{fig:simflux}
\end{figure}

Though it is difficult to infer a mass distribution of the HVCs and
CHVCs due to their unknown distances (M$_{HI} \propto$ D$^2$), we can
infer a \HI\ flux distribution of the halos using their
masses and distances and assuming  
a certain percentage of \HI\ gas is associated with the dark matter.  If the halos contain 1\% of their dark matter mass
in neutral hydrogen, their total HI fluxes are directly comparable to
those of the HVCs and CHVCs, as shown in Figure 5.  Note that some of the
plots have different values on both axes.
The projected \HI\ flux distribution of all of the dark matter halos is 
shown in the bottom left panel of Figure 5 by the solid points.  The slope is -1.5,
which is the same as that of the HVCs shown in the upper left panel, but the
number of halos in each flux bin is much higher.  The upper right panel shows
that both the total number of CHVCs and
the slope of the \HI\ flux distribution ($f(F_{HI}) \propto F_{HI}^{-1.8}$) is different from that
of the halos.  The selected populations of halos discussed above are shown
in the bottom two panels of Figure 5.  Halos outside of 300 kpc from the
Galaxy and M31 are represented by the dashed line in the bottom left panel.  The total number of clouds is again
somewhat higher than that of the HVCs and the slope is steeper, like the
CHVCs.  The \HI\ flux distribution of
those halos with M$_{DM} > 2 \times 10^8$ \Msun\ ($f(F_{HI}) \propto
F_{HI}^{ -1.55}$; solid points in the bottom right panel) and those halos within 500 
kpc of the Galaxy ($f(F_{HI}) \propto F_{HI}^{-1.6}$; dashed line in the
bottom right panel) are the most comparable to the HVCs in terms of total number
and slope.  
The size distribution of the halos is another property which
can be compared to the HVCs if one assumes that the gas and dark matter are 
coupled.  These plots will be shown in a future paper.  

Figure 5 also shows that even if the Local Group halos are not HVCs or
CHVCs, they should have been detected by the 
recent large area HI surveys if they contain $\sim$1\% of their dark 
matter mass in neutral hydrogen.
For instance, HIPASS covers over half of the sky and the velocities of the Local 
Group galaxies with a completeness
level of $\sim 4$ Jy \kms, and the Wakker catalog reaches fluxes of $\sim 5$
Jy \kms\ (Wakker \& van Woerden 1991).  Therefore, 
considering the entire population of halos (2135 total), almost all of them would have 
been detected if they contain 1\% or greater of their dark matter mass in \HI. 
By applying cuts in halo distance or mass, the number of halos which should have been detected is in closer agreement with the number 
of HVCs without lowering M$_{HI}$ to below  
1\% M$_{DM}$.  

\section{Discussion}

There does not appear to be a close correlation between the entire
population of dark matter halos in this standard cold dark matter Local
Group simulation and the high-velocity clouds and compact high-velocity
clouds.  However, when cuts are made in distance or mass, the remaining
halos show properties which could be argued are consistent with those of
the HVCs without distance constraints.  The compact HVCs do not show similar
properties to the dark matter halos.  The justification behind the halo cuts are
summarized below.  For greater than 300 kpc from M31 and the Galaxy, it is
possible that the neutral hydrogen gas would not survive outside of a
particular distance from these galaxies.  This could be due to either ram
pressure or tidal stripping effects (e.g. Quilis \& Moore 2001), or
ionizing radiation escaping from the galaxy (e.g. Bland-Hawthorn \& Maloney
1999). 
The cut which requires the halos associated with neutral hydrogen gas be
within 500 kpc, is based on the idea that the \HI\ is only able to condense
within a certain radius of the Galaxy (e.g. Oort 1970).  Finally, the
choice of
only those halos with dark matter masses above $2 \times 10^8$ \Msun\
having neutral gas associated with them comes from CDM simulations
which include reionization at $z \approx 6 - 10$ (e.g. Gnedin 2000).  The
increase in temperature of the gas during reionization suppresses the
formation of low mass satellites due to a reduction in gas mass via
the expulsion of already accreted gas and the suppression of further accretion.
The primary difference between these sub-populations of halos and the entire
population is a reduction in the total number, but there are
also differences in the flux distributions and within 500 kpc there is
certainly a more uniform spatial and kinematic distribution.

Considering the distance determinations which have been made to some of the
larger high velocity complexes and the search for HVCs in other groups
 which have not found extragalactic HI clouds without stars down to
M$_{HI}$ levels of $7 \times 10^6$ \Msun\ (e.g. Zwaan 2001), the possibility
of the HVCs being the dark matter halos within a certain radius of the
Milky Way seems the most likely possibility.
If the nearby halos are represented by HVCs, are the further away
halos represented by the lower column density Ly$\alpha$ absorbers which
have been detected in other groups (e.g. Penton, Shull \& Stocke 2000)?   It is
difficult to find these low column density systems within the Local Group due to the
the damping wings of the Galaxy's Ly$\alpha$ and to some extent Ly$\beta$
absorption lines
obscuring those velocities.  The detection of these systems through other
absorption lines may be possible in some
cases (e.g. $N_{HI} > 10^{14}$), and may have already been detected along some sightlines
(e.g. Sembach et al. 1999). Evidence for a link between HVCs and these few
absorption line detections of lower \HI\ column density high velocity gas lies
in their spatial and kinematic relationship.
Therefore, if this gas is related to the
Ly$\alpha$ absorbers, one might expect to detect higher column density
systems akin to HVCs in the
vicinity of some of the low-redshift Ly$\alpha$ absorber systems. 

 The \HI\ flux distribution plots of the halos, presented in the bottom two panels of
Figure 5, put strong constraints on the amount of neutral hydrogen the
halos can have.  HIPASS and other HVC surveys cover the velocities of all of
the Local Group galaxies down to \HI\ flux levels of 2-5 Jy \kms.  Figure 5
shows that the majority of the halos should have been detected if they
contain at least 1\% of their dark matter mass in \HI.  Since only
$\sim20$ Local Group galaxies have been detected in \HI, there is an
obvious need to look towards other objects.  The number of HVCs and CHVCs
is also too low to agree with the number of halos that should have been
detected, unless one applies distance and/or mass cuts (see Figure
5 and Table 1).  No other candidate ``missing satellites'' have
been detected in \HI\ in the Local Group, so either the HVCs are a
population of the missing satellites, or the majority of the dark matter
halos contain $< 1$\% of their dark matter mass in \HI.

\acknowledgements{We thank Bart Wakker for the use of his revised HVC
catalog, Lister Staveley-Smith and Ken Freeman for feedback on this work,
and the Multibeam Working Group and Vincent de Heij for help with the HIPASS data.}


\begin{references}
\reference{Barnes et al. 2001, \mnras, 322, 486}
\reference{Bland-Hawthorn, J. \& Maloney, P.R. 1999, ApJ, 510, L33}
\reference{Braun, R. \& Burton, W.B. 1999, A\&A, 341, 437} 
\reference{Freedman, W. \& Madore, B. 1990, \apj, 365, 186}
\reference{Gnedin, N. 2000, ApJ, 542, 535}
\reference{Hulsbosch, A.N.M. \& Wakker, B.P. 1988, A\&AS, 75, 191}
\reference{Moore, B. et al.
 2001, Phys. Rev. D. in press, (astro-ph/0106271)}
\reference{Morras, R., Bajaja, E., Arnal, E.M. \& Poppel, W.G.L. 2000, A\&AS, 142, 25}
\reference{Oort, J. H. 1970, A\&A, 7, 381}
\reference{Penton, S., Shull, M., \& Stocke, J. 2000, \apj, 544, 150}
\reference{Putman, M. et al. 2001, AJ accepted}
\reference{Putman, M. 2000, PhD Thesis, Australian National University}
\reference{Quilis, V. \& Moore, B. 2001, ApJ, 555, 95} 
\reference{Sembach, K. R., Savage, B. D., Lu, L., \& Murphy, E. 1999, \apj,
515, 108}
\reference{Wakker, B. 2001, ApJ, in press}
\reference{Wakker, B. \& van Woerden, H. 1991, A\&A, 250, 509} 
\reference{Zwaan, M. 2001, \mnras, 325, 1142}
\end{references}
\end{document}